\documentclass[aps,prl,twocolumn,superscriptaddress,showpacs,letterpaper]{revtex4-1}

\usepackage{graphicx}% Include figure files
\usepackage{dcolumn}% Align table columns on decimal point
\usepackage{bm}% bold math
\usepackage{amsmath}
\usepackage{multirow}

\begin{document}

\newcommand{\mevcc}{\!\mathrm{MeV}\!/c^2}
\newcommand{\mevc}{\!\mathrm{MeV}/\!c}
\newcommand{\mev}{\!\mathrm{MeV}}
\newcommand{\gevcc}{\!\mathrm{GeV}/\!c^2}
\newcommand{\gevc}{\!\mathrm{GeV}/\!c}
\newcommand{\gev}{\!\mathrm{GeV}}
\title{Is the Exotic Hadron X(3872) a $D^0\overline{D}^{*0}$ Molecule: Precision Determination of the Binding Energy of X(3872)}

\author{A.~Tomaradze}
\author{S.~Dobbs}
\author{T.~Xiao}
\author{Kamal~K.~Seth}
\affiliation{Northwestern University, Evanston, Illinois 60208, USA}

\author{G. Bonvicini} 
\affiliation{Wayne State University, Detroit, Michigan 48202, USA} 

%\date{July 14, 2009}
\date{\today}

\begin{abstract} 
It has been proposed that the recently discovered archetypical ``exotic'' meson, X(3872), with $M(\mathrm{X}(3872))=3871.68\pm0.17$~MeV/c$^{2}$, and an extremely narrow width, $\Gamma(\mathrm{X}(3872))<1.2$~MeV, is a hadronic molecule of bound $D^0$ and $\overline{D}^{*0}$ mesons. If true, this would establish a new species of hadrons, distinct from $q\bar{q}$ mesons and $qqq$ baryons. It is put to an important experimental test by making a high precision measurement of the proposed molecule's binding energy.  Using 818~pb$^{-1}$ of $e^+e^-$ annihilation data taken with the CLEO-c detector at $\psi(3770)$, the decays $D^0\to K_SK^+K^-$ and $D^0(\overline{D}^0)\to K^\pm\pi^\mp \pi^+\pi^-$ have been studied to make the highest precision measurement of $D^0$ mass, $M(D^0)=1864.851\pm0.020\pm0.019\pm0.054$~MeV/c$^{2}$, where the first error is statistical, the second error is systematic, and the third error is due to uncertainty in kaon masses, or $1864.851\pm0.061$~MeV/c$^{2}$ with all errors added in quadrature. This leads to $M(D^0+D^{*0})=3871.822\pm0.140$~MeV/c$^{2}$, and the binding energy $\mathrm{BE}(\mathrm{X}(3872))\equiv M(D^0+D^{*0})-M(\mathrm{X}(3872)) = +142\pm220$~keV.  At the 90\% confidence level this leads to the conclusion that X(3872) is either unbound by as much as 140~keV, or it is bound by less than 420~keV.  If bound, X(3872) has a very large radius; the central value of binding energy corresponds to a radius of 12~fm, and the lower limit to 7~fm, both being uncomfortably large for a molecule.
\end{abstract} 

\maketitle
%\large

%The unexpected narrow state, $X(3872)$, has become the poster--child 
%of charmonium--like exotic states. With $M(X(3872))=3871.68\pm 0.17$ MeV,
%and $\Gamma(X(3872))<1.2$ MeV \cite{pdg}, it does not fit into the expected
%charmonium spectrum, and 

Recent observations at the B--factories of many unexpected resonances, loosely called ``exotic'', have given rise to great excitement in the spectroscopy of heavy-quark hadrons.  In the mass region above bound charmonium resonances, $M>3.73$~GeV/c$^{2}$, most of the observed resonances have widths which range from 30 to 200~MeV, and many remain potential charmonium candidates. However, one resonance, dubbed X(3872), has acquired the status of the quinessential exotic because of its unique properties. It has been observed in many diverse experiments, by Belle~\cite{belleX,belleX2}, BaBar~\cite{babarX,babarX2}, CDF~\cite{cdf,cdf2}, D\O~\cite{d0X}, LHCb~\cite{lhcb}, and CMS~\cite{cms}. The closeness of its mass, $M(\mathrm{X}(3872))=3871.68\pm0.17$~MeV/c$^{2}$~\cite{pdg} to the sum of the masses of two open charm mesons $D^0(J^P=1^-)$ and $D^{*0}(J^P=1^-)$, and its extremely narrow width, $\Gamma(\mathrm{X}(3872))\le1.2$~MeV ($90\%$ CL), has given rise to the proposal that it is a $D^0\overline{D^{*0}}$ molecule. Many of its decays have been measured, and it is found that all decay final states invariably contain a charm quark and an anticharm quark, which would suggest that it is a narrow charmonium resonance. Angular correlation measurements limit its $J^{PC}$ to $1^{++}$ or $2^{-+}$~\cite{cdf3,belle}, so that the likely charmonium states would be $2^3P_1(1^{++})\chi'_{c1}$ or $1^1D_2(2^{-+})\eta_{c2}$. Unfortunately, the predicted masses of both these states are quite far ($\sim+75$~MeV/$c^2$, and $\sim-40$~MeV/$c^2$, respectively) from 3872~MeV/$c^2$, which makes it difficult to identify X(3872) as a pure charmonium state.

Numerous theoretical models for X(3872) have been proposed, and several reviews of the different possibilities exist in the literature~\cite{x-cc}. However, despite some problems, the most popular explanation for X(3872) remains that it is a loosely bound molecule of the $D^0$ and $D^{*0}$ mesons. If this explanation is correct, X(3872) would be member of a new species of hadrons, distinct from $q\bar{q}$ mesons and $qqq$ baryons. This indeed would be a most dramatic development in hadron spectroscopy, one that needs to be submitted to critical scrutiny.

Obviously, one of the most important properties of a molecule is its binding energy and it is necessary to make an accurate determination of it. Since the difference between the masses of the $D^0$ and $D^{*0}$ mesons has been accurately measured to be $142.12\pm0.07~\mathrm{MeV/c^2}$~\cite{pdg}, 
a precision determination of the 
binding energy of X(3872) as a $D^0\overline{D}^{*0}$ molecule
requires the highest precision measurement of the mass of the $D^0$ meson. 
In this letter we report on such a measurement.

We had earlier \cite{d0pub} reported the measurement of $M(D^0)$  
in the decay  $D^0\to K_S\phi$, $\phi\to K^+K^-$ using  280 pb$^{-1}$ of CLEO-c data 
taken at the $\psi(3770)$. 
 We determined
$M(D^0)=1864.847\pm0.178$ MeV/c$^{2}$, 
 which led to an uncertainty of $\pm363$~keV/c$^{2}$
in the mass $[M(D^0)+M(\overline{D}^{*0})]$.
 With the then known value of 
$M(\mathrm{X}(3872))$ which had an uncertainty of $\pm500$~keV/c$^{2}$, the 
binding energy was determined to be 
$\mathrm{BE(X(3872))}\equiv M(D^0+D^{*0})-M(\mathrm{X}(3872))=600\pm600$~keV/c$^{2}$. Since then, several 
improved measurements of $M(\mathrm{X}(3872))$ have been made \cite{belleX2,babarX2,cdf2,lhcb,cms}, with
the present PDG average   $M(\mathrm{X}(3872))=3871.68\pm 0.17$ MeV/c$^{2}$~\cite{pdg}, and it is now necessary to make a correspondingly more precise measurement of 
 $[M(D^0)+M(\overline{D}^{*0})]$ in order to determine $\mathrm{BE(X(3872))}$
with higher precision. In this letter, we report such a measurement,
which raises serious questions for the 
$|D^0\overline{D}^{*0}\rangle$ molecule model of X(3872).

A nearly factor three improvement in the precision of $M(D^0)$ has become 
possible because of two reasons. We now have nearly three times more CLEO-c 
data available, 
 $\sim818$~pb$^{-1}$ of data 
taken at the $\psi(3770)$, $\sqrt{s}=3770$~MeV,  % , and 
% $\sim586$~pb$^{-1}$ of data 
%taken at the $\sqrt{s}=4170$~MeV, 
and in addition to  $D^0\to K_SK^+K^-$,
we study the nearly forty times more prolific 
decay, $D^0(\overline{D}^0)\to K^{\pm}\pi^{\mp}\pi^{+}\pi^{-}$. The data taken at $\sqrt{s}=3770$~MeV is ideally suited for this measurement because $\psi(3770)$ decays almost exclusively to $D\overline{D}$ ($\mathrm{branching~fraction}=93^{+8}_{-9}\%$), and the $D$--mesons are produced almost at rest.

In the present investigation we determine the mass of $D^0$ with a precision
of $\sim60$~keV/c$^{2}$. This requires improvement of the default 
solenoid magnetic field calibration of CLEO-c, and to track its small variation with time. We do the calibration by choosing
to anchor our mass measurements to the high precision 
measurements of the masses of $\psi(2S)$ and $J/\psi$ with uncertainties of $\pm15$~keV/c$^{2}$ and $\pm12$~keV/c$^{2}$, respectively, made by the KEDR Collaboration by the 
resonance--depolarization technique~\cite{kedr,kedr1}.
Our investigation involves several steps. We first recalibrate the CLEO-c 
solenoid magnetic field using the KEDR masses in a study of the exclusive 
decay  $\psi(2S)\to \pi^{+}\pi^{-} J/\psi$ using CLEO-c data for  
25 million $\psi(2S)$. With the recalibrated  magnetic field we make 
a precision measurement of the mass of $K_S$ in the inclusive decay,
$\psi(2S)\to K_S+X$. Using $M(K_S)$ so determined we do fine tuning of the 
 magnetic field for each individual CLEO-c dataset at  
$\sqrt{s}=3770$~MeV via the inclusive decay
$D\to K_S+X$. We use these fine tuned fields to make our measurements 
of $D^0$ mass in the two exclusive decays:  $D^0\to K_SK^+K^-$
and $D^0(\overline{D}^0)\to K^{\pm}\pi^{\mp}\pi^{+}\pi^{-}$.

The data were taken with CLEO-c detector \cite{CLEOcDetector}, which consists of a CsI(Tl) electromagnetic calorimeter, an inner vertex drift chamber, a central drift chamber, and a ring imaging Cherenkov (RICH) detector, all inside a superconducting solenoid magnet providing a 1.0 Tesla magnetic field.  For the present measurements, the important components are the drift chambers, which provide a coverage of 93\% of $4\pi$ for the charged particles. 
 The detector response was studied using a GEANT-based Monte Carlo (MC) simulation including radiation corrections~\cite{GEANTMC}. 

For the analysis of $\psi(2S)$ decays, $\psi(2S) \to \pi^{+}\pi^{-} J/\psi$, 
$J/\psi \to \mu^{+}\mu^{-}$, and $\psi(2S)\to K_S+X$,  
$K_S\to \pi^{+}\pi^{-}$,  we select events with
 well-measured tracks by requiring that they be fully contained in the 
barrel region ($|\cos\theta| < 0.8$) of the detector, and have transverse
 momenta $>120$~MeV/$c$.
 For the pions from $K_S$ decay, we make the additional requirement that 
they originate from a  common vertex displaced from the interaction point 
by more than 10 mm. We require a $K_S$ flight distance significance of more 
than 3 standard deviations.
 We accept $K_S$ candidates with mass in the range $497.7\pm12.0$ 
MeV/c$^{2}$.  
We identify muons from $J/\psi$ decays as having 
momenta more than 1 GeV, and $E_{CC}/p<0.25$ for at least one muon candidate,
and  $E_{CC}/p<$0.5 for the other muon. We require that there should be only two identified 
pions and two identified muons with opposite charges in the event.
The momenta of $\mu^{+}\mu^{-}$ pairs is kinematically fitted to the 
KEDR $J/\psi$ mass, $M(J/\psi)=3096.917$~MeV/c$^{2}$,  
and only events with $\chi^2<20$ are
accepted. We also require that there should not be any isolated shower 
with energy more than 50 MeV in the event.

As stated earlier, to make a precision recalibration of the solenoid 
magnetic field we reconstruct $\psi(2S)$ in the decay  
$\psi(2S)\to \pi^{+}\pi^{-} J/\psi$, $J/\psi\to \mu^{+}\mu^{-}$.
Using KEDR masses, $M(\psi(2S))_\mathrm{KEDR}=3686.114\pm0.015$~MeV/$c^2$~\cite{kedr}
and $M(J/\psi)_\mathrm{KEDR}=3096.917\pm0.012$~MeV/$c^2$~\cite{kedr1}, we
 determine the magnetic field correction required to modify the pion momenta such that 
$M(\psi(2S))_\mathrm{Present}$ becomes identical to 
$M(\psi(2S))_\mathrm{KEDR}$.
With pions with momenta $<600$~MeV/$c$, the required correction is determined
to be $+0.029\%$ (or~0.29~Gauss) in the default CLEO calibration of the magnetic
field. Fig. 1(top) shows the distribution of 
$\Delta M(\psi(2S))\equiv M(\psi(2S))_\mathrm{Present})-M(\psi(2S))_\mathrm{KEDR}$ 
after the correction. 
We make fits to the unbinned data in the full range with peaks parameterized as sum of a simple
Gaussian function and a bifurcated Gaussian function, and a linear background.
The fit has the number of events,  $N(\psi(2S))=125300\pm356$, and $\chi^2$/dof=1.00, and gives
 $\Delta M(\psi(2S))=0.0\pm6.7$ keV/c$^{2}$.
An identical procedure is used to fit all other mass distributions presented 
in this letter.

Having corrected the magnetic field, we use it to analyze the same
$\psi(2S)$ data set for the inclusive decay, $\psi(2S)\to K_S+X$,  
$K_S\to \pi^{+}\pi^{-}$ for pions in the same momentum region, $<600$~MeV/c$^{2}$.
The fit to the $\pi^{+}\pi^{-}$ invariant mass distribution, shown in 
Fig. 1 (bottom), has  $\chi^2$/dof=1.06. It leads to the number of $K_S$, 
$N(K_S)=256859\pm739$, and
\begin{equation}
M(K_S)_{\mathrm{Present}}=497.600\pm0.007(\mathrm{stat})\;\mathrm{MeV/c^{2}}.
\end{equation}

\begin{figure}[!t]
%\vspace*{-1.cm}
\includegraphics*[width=3.4in]{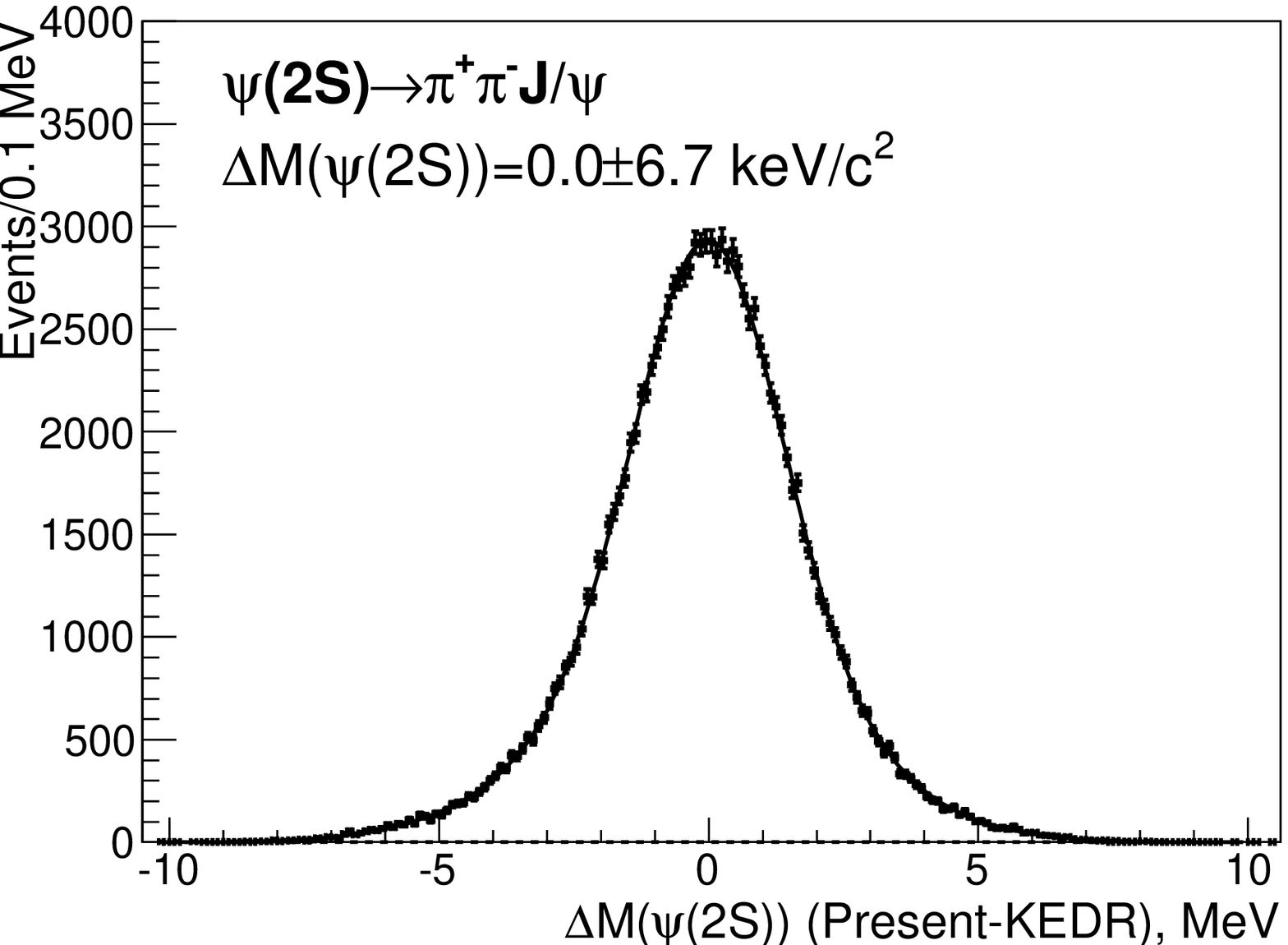}
\includegraphics*[width=3.4in]{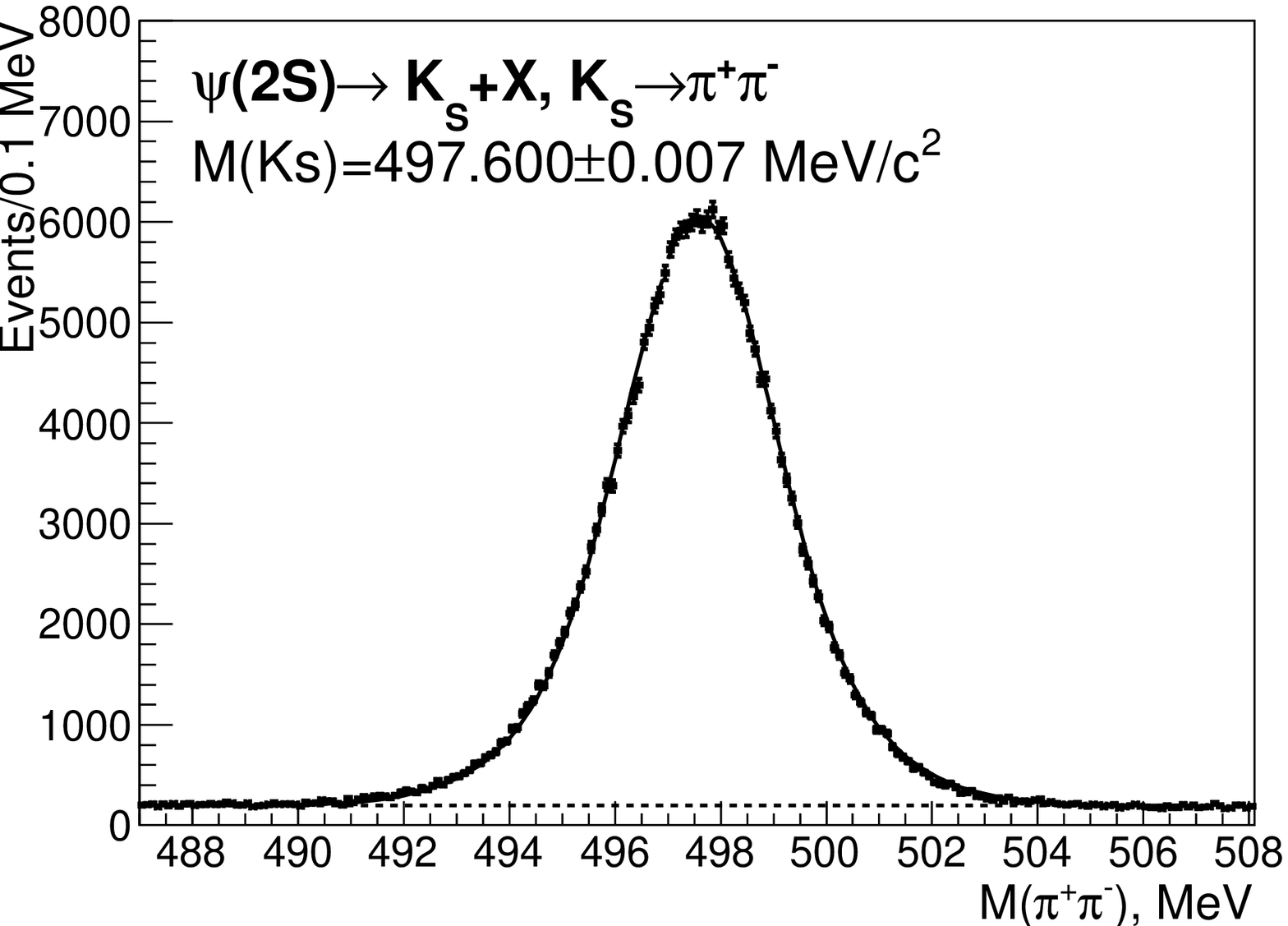}
\caption{Top plot: 
$\Delta M(\psi(2S))\equiv M(\pi^{+}\pi^{-}J/\psi)-M(\psi(2S))_{KEDR}$
for the exclusive decays $\psi(2S) \to \pi^{+}\pi^{-} J/\psi$, 
$J/\psi \to \mu^{+}\mu^{-}$ using the corrected magnetic field. 
 Bottom plot:  $M(\pi^+\pi^-)$ for the inclusive reaction
$\psi(2S)\to K_S+X$,  $K_S\to \pi^{+}\pi^{-}$, using the corrected 
magnetic field.
The curves show peak fits with  a sum of a simple
Gaussian function and a bifurcated Gaussian function and a linear background.}
\end{figure}
\begin{figure}
\begin{center}
\includegraphics[width=3.4in]{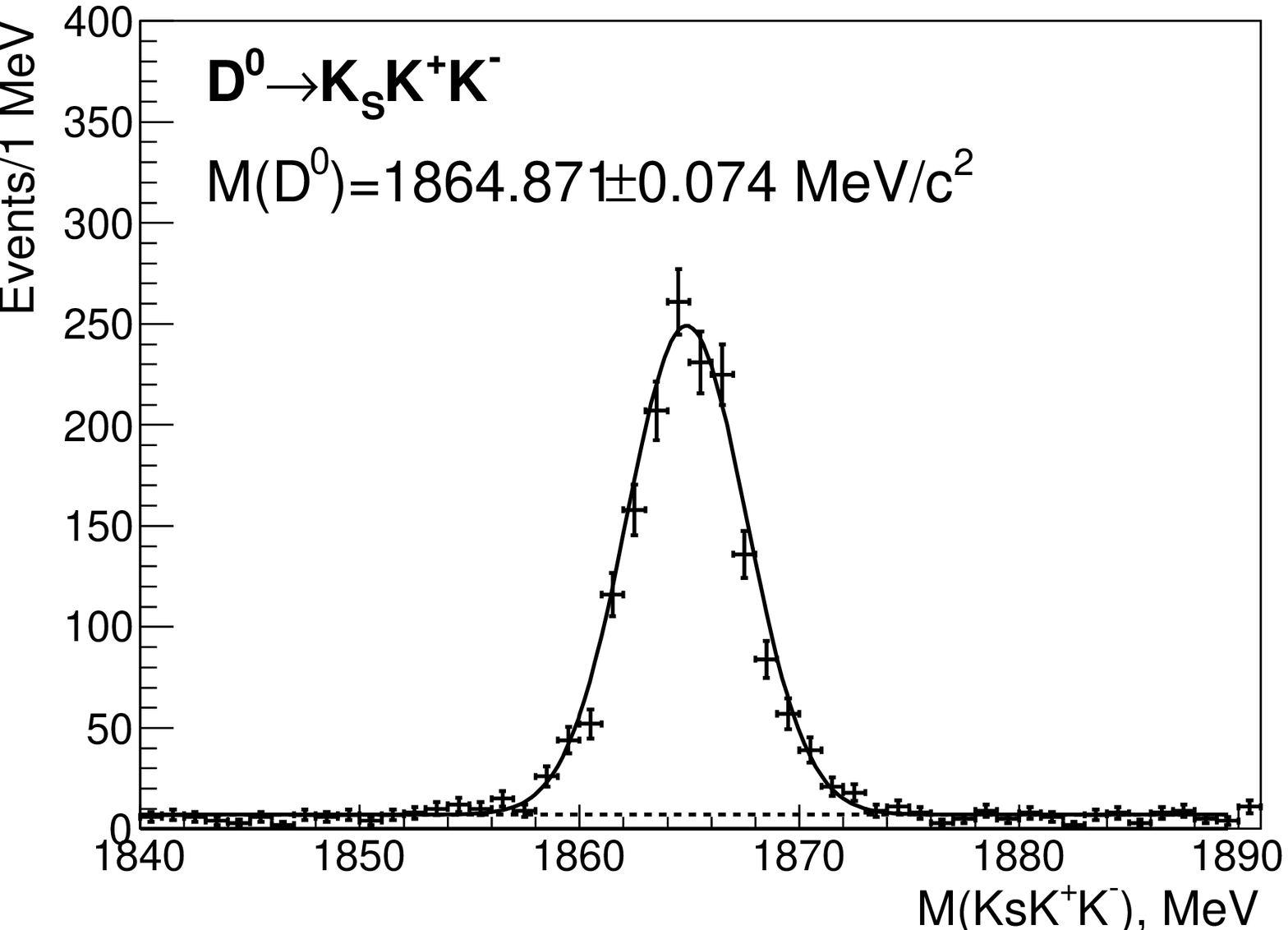}
\includegraphics[width=3.4in]{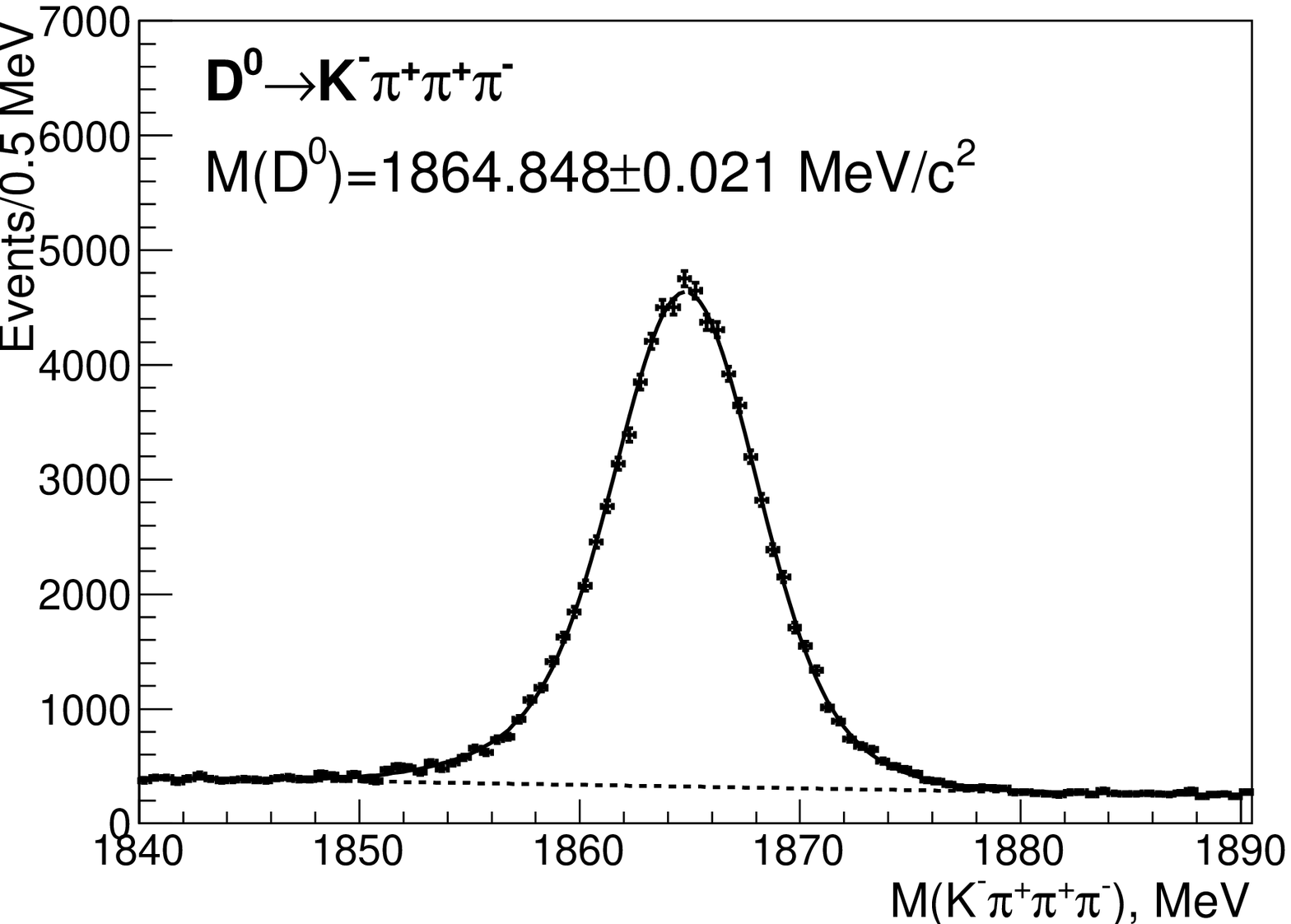}
\end{center}
\caption{Invariant mass spectra  for the decays
(top) $D^0\to K_SK^+K^-$,
(bottom) $D^0\to K^{\pm}\pi^{\mp}\pi^{+}\pi^{-}$.
Fitted masses with statistical errors only are given.}
\end{figure}

We next select $D^0$ candidates in the data taken at $\sqrt{s}=3770~\mathrm{MeV}$ using the standard CLEO D--tagging criteria, 
which impose a very loose requirement on the beam energy constrained 
$D^0$ mass, as described in Ref. \cite{dtag}.  
%In order to correct for the possible small variations of the magnetic
%field in the different data sets taken at both  $\sqrt{s}$=3770 MeV and
% $\sqrt{s}$=4170 MeV the inclusive decay, $D\to K_S+X$ was measured for 
%each individual data set, and corrections to the default CLEO-c magnetic 
%field were determined to bring each $M(K_S)$ measured to became equal 
%to $M(K_S)_{CLEO}$ in Eq. 1. 
Because the data at $\sqrt{s}=3770$~MeV were taken in several smaller sub--runs, the solenoid magnetic field needs to be corrected for possible small variations from sub--run to sub--run.  We do so by analyzing each sub--run for the inclusive decay $D\to K_s +X$, $K_S\to\pi^+\pi^-$ and requiring that the field be corrected to bring each $M(K_S)$ to the value $M(K_S)_\mathrm{CLEO}$ in Eq.~1.
These corrections were found to be $<\pm0.030\%$, 
consistent with what was found for $\psi(2S)$. With these corrections 
in place, individual data sets were analyzed for the decays 
$D^0 \to  K_S K^+K^-$, $K_S\to\pi^+\pi^-$, and
$D^0(\overline{D}^0) \to  K^{\pm}\pi^{\mp}\pi^{+}\pi^{-}$.
In order to use the same magnetic field calibration 
for all decay channels, we use only those events in
which final state pions and kaons have momenta  $<600$~MeV/$c$. 
We select well-measured tracks which  
have specific ionization energy loss, $dE/dx$, 
in the drift chamber consistent with pion or kaon hypothesis within 3 
standard deviations. 
For the $K_S$ candidates from the exclusive $D^0$   
decays, we perform a mass-constrained (1C) kinematic fit, and accept in our 
final sample $K_S$ with $\chi^2<20$.
%We select events containing a $\phi$ by requiring that $M(K^+K^-)$ of the candidate kaons is within $\pm15~\mathrm{MeV}$ of the 
%value $M(\phi)=1019.46$~MeV \cite{pdg}. 

%The mass distributions for summed data for for each decay were fitted as
%described. The fits for the two decay channels are shown in Fig. 2 (top) 
%for data at  $\sqrt{s}$=3770 MeV,
%and in Fig. 2 (bottom) for data at  $\sqrt{s}$=4170 MeV. 
%The  $M(D^0)$ values in data were corrected on the Monte-Carlo 
%output--input mass differences, with the output 
%$D^{0}$ mass values obtained using the same procedure as in the
%data. The results are shown in Table I.
%The values for  $M(D^{0})$ agree for the 3770 MeV and 4170 MeV data,
%and for the different decay modes.

The final mass distributions for the different sub--runs were added together, and total data at $\sqrt{s}=3770$~MeV  for the decays, $D^0\to K_SK^+K^-$ (which includes events in which $K^+K^-$ form a $\phi$ meson) and $D^0(\overline{D}^0)\to K^\pm\pi^\mp\pi^+\pi^-$ ($K3\pi$ henceforth) are shown in Fig.~2.  The distributions were fitted with peaks as described before, and linear backgrounds.  The fits have 
full widths at half maximum of 6.4~MeV for $K_SK^+K^-$, and 8.1~MeV for $K3\pi$, and  $\chi^2$/dof of 
1.02 and 1.04, respectively.  The results of the fits are listed in Table~I.  Both results for $M(D^0)$ 
are seen to be consistent within statistical errors, and their average is 
\begin{equation}
M(D^0)_{Present}=1864.851\pm0.020(\mathrm{stat})~\mathrm{MeV/c^{2}}.
\end{equation}
\begin{table}[htb]
\caption{Results of fits to the mass distributions.  The systematic errors are as listed in Table~II and described in the text.}

%\large
%\begin{small}
%\begin{center}
\begin{ruledtabular}
\begin{tabular}{lcc}
%\hline
Decay & $N$(events) & $M(D^0)$, MeV/c$^{2}$ \\
\hline

$D^0\to K_SK^+K^-$ & $1,655\pm43$  & $1864.871\pm0.074\pm0.063$ \\
$D^0\to K3\pi$   & $76,988\pm388$ & $1864.848\pm0.021\pm0.061$ \\
\hline
Average          &     ---        & $1864.851\pm0.020\pm0.057$ \\
%$D^0$ decay mode & $N_{event}$  & resolution $\sigma$, MeV &  MC(out-inp), keV & $M(D^0)$, MeV  \\
%\hline
%$K_{S}\phi$ at 3770 MeV & 883$\pm$30  &2.65$\pm$0.07 & 4$\pm$ 7 & 1864.806$\pm$0.094  \\
%$K_{S}\phi$ at 4170 MeV & 1036$\pm$41  &2.70$\pm$0.09 & 15$\pm$8 & 1864.730$\pm$0.091  \\
%$K_{S}\phi$ all         &1919$\pm$51  &2.67$\pm$0.06 & 9$\pm$ 5 & 1864.767$\pm$0.065$\pm$0.052  \\
% &  & &  &  \\
%$K^{\pm}\pi^{\mp}\pi^{+}\pi^{-}$ at 3770 MeV &71988$\pm$388  &3.45$\pm$0.06 & 7$\pm$ 4 & 1864.841$\pm$0.021  \\
%$K^{\pm}\pi^{\mp}\pi^{+}\pi^{-}$ at 4170 MeV &50502$\pm$601  &3.39$\pm$0.08 & 2$\pm$11 & 1864.796$\pm$0.033  \\
%$K^{\pm}\pi^{\mp}\pi^{+}\pi^{-}$ all &122490$\pm$715  &3.42$\pm$0.05 & 6$\pm$ 4 & 1864.828$\pm$0.018$\pm$0.059  \\
% &  & &  &  \\
%both modes total  &124409$\pm$716  &3.11$\pm$0.04 & 7$\pm$ 3 & 1864.821$\pm$0.017$\pm$0.057  \\

%\hline
\end{tabular}
\end{ruledtabular}
%\end{center}
%\end{small}
\end{table}

% The corrections were found to be --26$\pm$6 keV, --31$\pm$5 keV and 
%+52$\pm$8 keV for  decays   $D^0\to K_S \phi$,
%$D^0\to K^{\pm}\pi^{\mp}\pi^{+}\pi^{-}$, 
%and $D^0\to K_S\pi^{+}\pi^{-}$, respectively.
%The values for  $M(D^{0})$ agree for the 3770 MeV and 4170 MeV data.
%The average values are: $M(D^0)=1864.692\pm0.082$ ($K_{S}\phi$ mode),
% $M(D^0)=1864.801\pm0.026$ ($K 3\pi$ mode),
% $M(D^0)=1864.839\pm0.082$ ($K_{S}2\pi$ mode).

The systematic errors in $M(K_S)$ and $M(D^0)$ were obtained as follows.

For $M(K_S)$ measurement, we have corrected the magnetic field using KEDR measured $M(\psi(2S))$ and $M(J/\psi)$, which have the total errors 
of $\pm15$~keV/$c^2$ and $\pm12$~keV/$c^2$, respectively~\cite{kedr,kedr1}. The change in $M(K_S)$ due to the change in
magnetic field is factor 1.46 smaller
than the change in $M(\psi(2S))$.
We therefore assign ($\pm15$/1.46)$\sim\pm$10.3 keV/$c^2$, and ($\pm12/1.46$)$\sim\pm8.2$~keV/$c^2$, as the uncertainties in $M(K_S)$ due to the uncertainties in  $M(\psi(2S))$ and $M(J/\psi)$.
%In addition to the KEDR value of $\psi(2S)$ mass, 
%we used the KEDR measured value of 
%$M(J/\psi)=3096.917\pm0.010\pm0.007$ MeV in the mass
%constrained fit of  $\mu^{+}\mu^{-}$ pairs for the
%exclusive sample $\psi(2S) \to \pi^{+}\pi^{-} J/\psi$, 
%$J/\psi \to \mu^{+}\mu^{-}$.
%The  $\pm$12 keV total uncertainty in $J/\psi$ mass yields  
%$\pm$12 keV uncertainty in  $\psi(2S)$ mass, and thus
%($\pm12$/1.46)$\sim\pm$8.2 keV
%uncertainty in  $K_S$ mass which we assign as systematic error.
The variation of the fit range by $\pm$2 MeV/$c^2$ yielded $\pm$4 keV/$c^2$ 
systematic error.
Changing the fits to the background from polynomials of order one to polynomials of order two changes $M(K_S)$ by $<1$~keV/$c^2$. 
%We assign 
%a systematic error of  $\pm1~\mathrm{keV}$   
%due to the background shape.
The effect of the possible formation of $\psi(2S)$ at an energy different
from the peak was investigated in detail.  
%The possible shift of produced $\psi(2S)$ mass was estimated 
This shift was estimated to be $\pm7$~keV/$c^2$, and it contributes $\pm5$~keV/$c^2$ to the
systematic error in $K_S$ mass. The sum in quadrature of all the above contributions is a total systematic 
uncertainty of $\pm15~\mathrm{keV/c^2}$ in $M(K_S)$. 

Our final result for $M(K_S)$ is thus
\begin{equation}
M(K_S)_{Present}=497.600\pm0.007\pm0.015\;\mathrm{MeV/c^2}.
\end{equation} 
Here, and elsewhere when mentioned separately, the first error is statistical and the second error is systematic. With statistical and systematic errors added in quadrature our result $M(K_S)_\mathrm{Present}=497.600\pm0.017$~MeV/$c^2$, is the world's most precise single measurement of $M(K_S)$, as illustrated in Fig.~3(top).  The PDG~2012 average of all previous measurements is $M(K_S)=497.614\pm0.022$~MeV/$c^2$.

\begin{table}[htb]
\caption{Systematic errors in $M(D^0)$ for range of variation of different parameters. The two values of the total 
systematic errors correspond to the uncorrelated systematic
errors, and the correlated systematic errors due to uncertainties in $K_S$ 
and $K^{\pm}$ mass measurements.}
%\large
%\begin{small}
%\begin{center}
\begin{ruledtabular}
\begin{tabular}{lcc}
%\hline
Source    & $\Delta M(D^{0})_{\mathrm{syst}}$   &  $\Delta M(D^{0})_{\mathrm{syst}}$   \\
 Systematic Error (keV/c$^{2}$) & $D^0\to K_S K^+K^-$  & $D^0\to  K3\pi$   \\
\hline
%Error in $K_S$ mass   & 47  & 55   \\
%Error in $K^{\pm}$ mass        & 22   & 13  \\
%Fit Range     & 3  & 15    \\ 
$|\cos\theta|_{\mathrm{max}}$: 0.8, 0.75    & 23  & 12    \\
$p_{\mathrm{min}}$(trans): 120, 135~MeV/c & 26  & 8    \\
$p_{\mathrm{max}}$(total): 650, 570~MeV/c  & 4  & 7    \\
Fit Range $\pm10$~MeV   & 3  & 15    \\ 
Bkgd.~Polynom. (1,2 order)  & 1  & 1  \\
MC Input/Output    & 5 & 5  \\
%\hline
Total (uncorrelated) & 35   & 22 \\
\hline\hline
Error in $K_S$ mass   & 47  & 55   \\
Error in $K^{\pm}$ mass        & 22   & 13  \\
Total (correlated)& 52   & 57 \\
\end{tabular}
\end{ruledtabular}
%\end{center}
%\end{small}
\end{table}

\begin{figure}[!t]
%\vspace*{-1.cm}
\includegraphics*[width=3.3in]{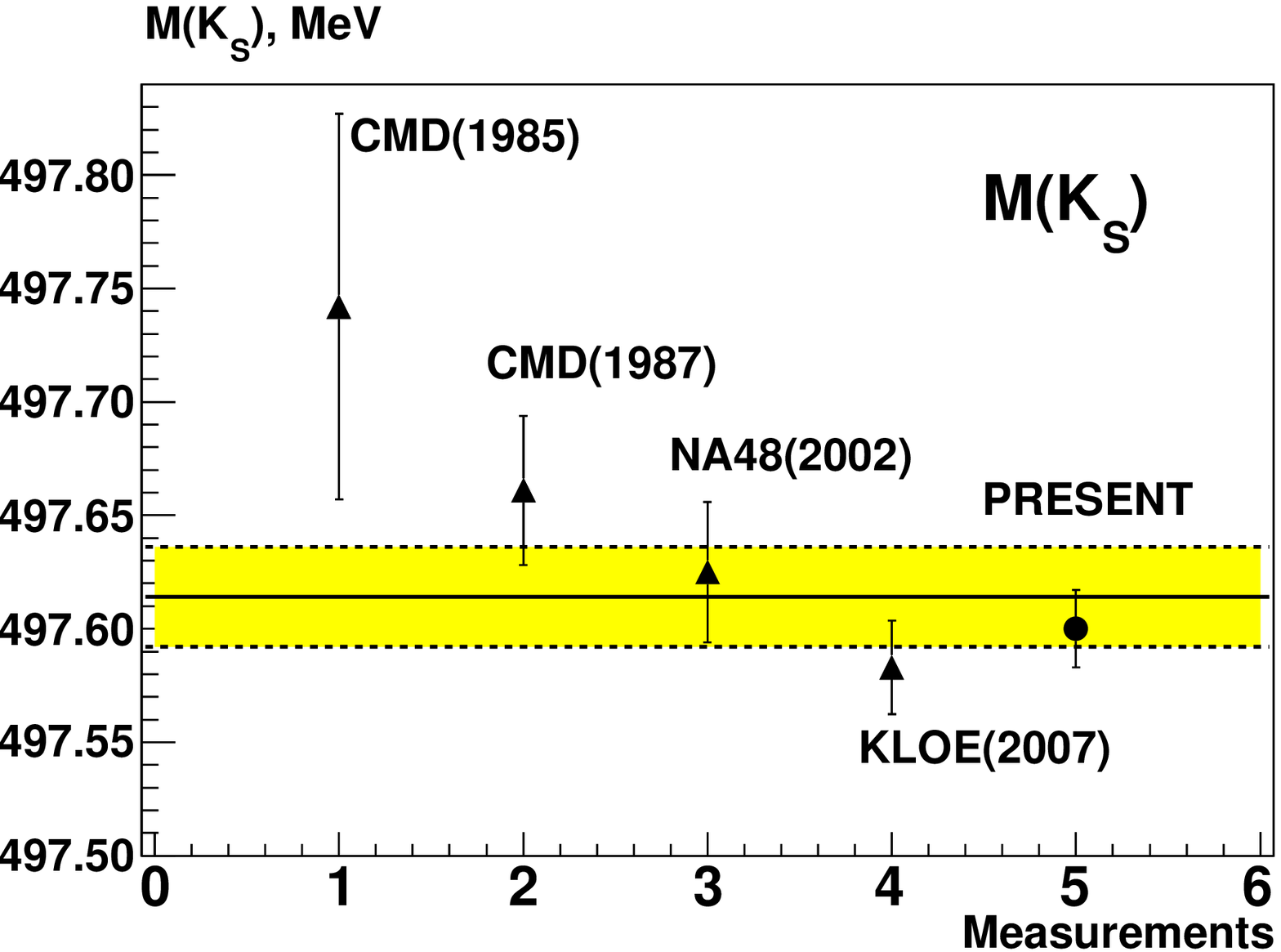}

\includegraphics*[width=3.3in]{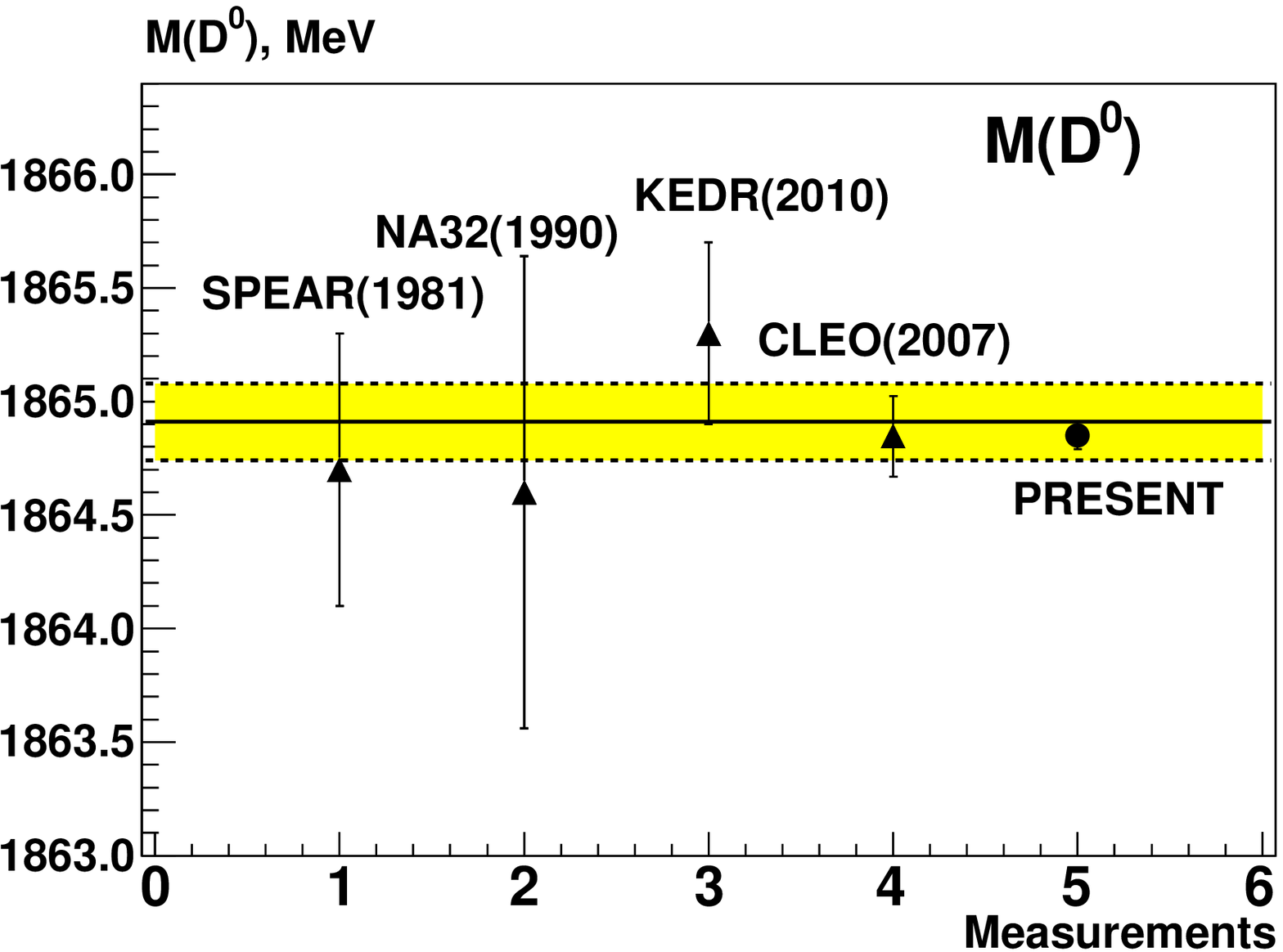}
\caption{Present results and previous mass measurements~\cite{pdg}, $M(K_S)$ (top) and $M(D^0)$ (bottom). The horizontal solid lines show PDG(2012) average.  The dashed lines show PDG(2012) error bands~\cite{pdg}.}
\end{figure}

The systematic errors in $M(D^0)$, as determined by varying event selection and peak fitting parameters, are summarized in Table~II. 
% The largest contribution comes from the $\pm17$~keV error in  our 
%measured $M(K_{S})$ in Eq. 2.
%The corresponding changes in $M(D^0)$ are $\pm47$~keV and $\pm55$~keV.
%The PDG(2012) mass $M(K^{\pm})=493.677$~MeV has an error of $\pm13$~keV~\cite{pdg}.
%This yields  $\pm$22 keV and $\pm$13 keV uncertainties in $M(D^0)$  for the decay modes of  
%$D^0\to K_S K^+K^-$ and $D^0\to K3\pi$, respectively.
Part of each $\Delta M(D^0)_{\mathrm{syst}}$ listed in the table can be due to changes in statistics when the parameter values are changed, but to be conservative we assign the full variations in $M(D^0)$ to the systematic errors. The correlated systematic errors listed in Table~II arise from uncertainties in the masses of $K_S$ and $K^\pm$. The $\pm17~\mathrm{MeV/c^2}$ uncertainty in the mass of $K_S$ used to fine-tune the solenoid magnetic field leads to the largest uncertainties in $M(D^0)$, $\pm47$~keV/$c^2$ in $D^0\to K_S K^+K^-$ and $\pm55$~keV/$c^2$ in $D^0\to K3\pi$. 
The PDG(2012) mass of $K^\pm$ has an error of 
$\pm13$~keV/$c^2$~\cite{pdg}. It leads to $\pm$22 keV/$c^2$ and $\pm$13 keV/$c^2$ uncertainties in $M(D^0)$ for the decay modes of $D^0\to K_S K^+K^-$ and $D^0\to K3\pi$.
%Thus, adding these contributions in quadrature yield  $\pm$52 keV and 
%$\pm$57 keV uncertainties in $M(D^0)$ for these decay modes.

In the two decay modes the sums in quadrature of the uncorrelated systematic 
uncertainties are 35~keV/$c^2$ and 22~keV/$c^2$, and the sum of correlated uncertainties due to 
$K_S$ and $K^{\pm}$ masses are 52~keV/$c^2$ and 57~keV/$c^2$. Taking proper account these uncertainties,
the average $D^0$ mass for the two decay modes is:
\begin{equation}
M(D^0)_{Present}=1864.851\pm0.020\pm0.019~\mathrm{MeV/c^2}.
\end{equation}
There is an additional uncertainty of $\pm$0.054~MeV/$c^2$ due to uncertainty in kaon masses. With all uncertainties added in quadrature, our present
result 
\begin{equation}
M(D^0)_{Present}=1864.851\pm0.061~\mathrm{MeV/c^2} 
\end{equation}
is the world's most precise single measurement of the mass of the $D^0$ meson, as illustrated in Fig.~3(bottom).
It supercedes our previous result in Ref.~\cite{d0pub} which was based on part of the data used in the present investigation. Our result for $M(D^0)$, and the PDG value 
$\Delta[M(D^{*0})-M(D^0)]=142.12\pm0.07$ MeV/$c^2$ \cite{pdg} lead to  
\begin{equation}
M(D^0+\overline{D}^{*0}) = 3871.822\pm0.140~\mathrm{MeV}/c^2,~~\mathrm{and} 
\end{equation}
\begin{equation}
\mathrm{BE}(\mathrm{X}(3872)) = +142\pm220~\mathrm{keV},
\end{equation}
using $M(\mathrm{X}(3872))=3871.68\pm0.17$ MeV/$c^2$ \cite{pdg}.  %, we obtain
%$B.E.(X(3872))= M(D^{0}+D^{*0})-M(X(3872))=+82\pm220$~keV.
At 90\% confidence level this result corresponds to X(3872) being
unbound by 140~keV, or being bound by at most 424~keV.  

As is well known, a universal property of a weakly bound system of two constituents with reduced mass $\mu$, and binding energy BE is that the root-mean-square separation of the constituents, or the ``radius'' of the composite, is given by $d=1/\sqrt{2\mu \mathrm{BE}}$. The central value and the $90\%$ CL upper limit of binding energy lead to
\begin{equation}
\begin{array}{l}
\displaystyle \mathrm{BE(X(3872))}=142~\mathrm{keV},~d\mathrm{(X(3872))}=12~\mathrm{fm} \\
\displaystyle \mathrm{BE}_{\mathrm{max}}(\mathrm{X}(3872))=424~\mathrm{keV},~d_{\mathrm{min}}(\mathrm{X(3872))}=7~\mathrm{fm} 
\end{array}
\end{equation}

With the early binding energy estimates of the order of 1~MeV~\cite{d0pub}, which 
corresponds to $d(\mathrm{X}(3872))=4.5$~fm, the long-range interaction responsible for the binding of X(3872) was suggested to be pion exchange. The present determination of $\mathrm{BE}=142\pm220$~keV corresponds to a radius as large as 12~fm, or at least 7~fm (90\% C.L.), twice as large as the deuteron, and it is difficult to see how pion exchange could explain the binding of $D^0$ and $D^{*0}$ into a molecule of this size. We recall that several other observations have also raised questions for the molecular model. These include too large a cross section for X(3872) formation at the Tevatron, too large a ratio $\sigma(\mathrm{X(3872)}\rightarrow\gamma\psi(2S))/\sigma(\mathrm{X(3872)}\rightarrow\gamma J/\psi$), and the possibility that $\mathrm{J^{PC}(X(3872))=2^{-+}}$~\cite{x-cc}. Together with our measurement of the uncomfortably large size, the explanation of X(3872) as a $D^0\overline{D}^{*0}$ molecule appears to have serious problems.
  %The $|D^0D^{*0}\rangle$ molecular model is seriously challenged.
%In summary, our measurement of the binding energy poses a serious challenge to the $|D^0\overline{D}^{*0}\rangle$ molecular model of X(3872).

%Among the universal properties of a molecule is the root-mean-square (rms) separation of its constituents, $d=1/\sqrt{2\mu (B.E.)}$, where $\mu$ is the reduced mass. With this binding energy measured by us the $D^0\overline{D}^{*0}$ molecule has a radius of $d>7.5$~fermi (90\% C.L.). It is interesting to note that this is almost twice the 4.3~fermi  radius of the deuteron.
 
%To summarize, we have determined the masses of the $K_S$ and $D^0$ mesons with the highest precision ever as shown in Fig. 3. We have determined the binding energy of the $X(3872)$ as the proposed   $D^0\overline{D}^{*0}$ molecule to be so small that within errors of the measurement it is either unbound or bound so weakly that its radius is more than twice as large as that of the  deuteron.

%\vspace*{2.cm}

This investigation was done using CLEO data, and as members of 
the former CLEO Collaboration we thank it for this privilege.
This research was supported by the U.S. Department of Energy.

\end{document}